\documentclass[superscriptaddress,floatfix,twocolumn]{revtex4-2}
\usepackage{times,fancyhdr}
\usepackage[dvips]{graphicx}
\usepackage{orcidlink}
\usepackage{amsmath,amssymb,bm,dsfont}
\usepackage{natbib}
\usepackage{color}
\usepackage{mathrsfs}
\usepackage{setspace}
\usepackage{hyperref}

\begin{document}
\title{Comment on ``Why interference phenomena do not capture the essence of quantum theory''}
\author{Jonte R. Hance\,\orcidlink{0000-0001-8587-7618}}
\email{jonte.hance@newcastle.ac.uk}
\affiliation{School of Computing, Newcastle University, 1 Science Square, Newcastle upon Tyne, NE4 5TG, UK}
\affiliation{Quantum Engineering Technology Labs, Department of Electrical and Electronic Engineering, University of Bristol, Woodland Road, Bristol, BS8 1US, UK}
\author{Sabine Hossenfelder\,\orcidlink{0000-0002-2515-3842}}
\affiliation{Munich Center for Mathematical Philosophy, Ludwig-Maximilians-Universit\"at M\"unchen, Geschwister-Scholl-Platz 1, D-80539 Munich, Germany}
\date{\today}

\begin{abstract}
 It was recently argued by Catani \textit{et al} that it is possible to reproduce the phenomenology of quantum interference classically, by the double-slit experiment with a deterministic, local, and classical model (\href{https://quantum-journal.org/papers/q-2023-09-25-1119/}{Quantum \textbf{7}, 1119 (2023)}). The stated aim of their argument is to falsify the claim made by Feynman (in his third book of Lectures on Physics) that quantum interference is ``impossible, {\textit{absolutely}} impossible, to explain in any classical way'' and that it ``contains the {\textit{only}} mystery'' of quantum mechanics. We here want to point out some problems with their argument.
\end{abstract}

\maketitle

\section{Introduction}

Catani \textit{et al} recently claimed to be able to reproduce the phenomenology of interference classically, by reproducing the double-slit experiment with a deterministic, local, and classical model that respects measurement independence (is not superdeterministic) \cite{Catani2021Interference}. They state that the aim of their argument is to falsify Feynman's claim that quantum interference is ``impossible, {\textit{absolutely}} impossible, to explain in any classical way'' and that it ``contains the {\textit{only}} mystery'' of quantum mechanics \cite{Feynman2011Vol3}. In this brief comment, we summarise how their argument works, then highlight some problems with it, and defend Feynman's position. Note, for the avoidance of doubt, this comment does not attempt to prove Feynman's claims, just illustrate issues with Catani \textit{et al}'s argument; to prove Feynman's claims would be an entire research programme of its own, but initial work in this direction has been undertaken in \cite{ji2024tracing}.

Note also that Catani \textit{et al} have previously attempted to respond to some of the problems we raised in previous versions of this comment, both in their paper \cite{Catani2021Interference}, and in a dedicated reply \cite{catani2022reply}.
Our comment has been revised to reflect their responses. Further, despite their paper \cite{Catani2021Interference} originally claiming that the entire phenomenology of interference can be described classically using their toy model, Catani \textit{et al} have since published a paper claiming to show ``aspects of the phenomenology of interference that are genuinely nonclassical'' \cite{Catani2023Nonclassical}, and so have reduced the scope of their claims in \cite{Catani2021Interference}.

\section{Catani \textit{et al}'s Argument}
 
Catani \textit{et al} present their argument in two steps. First, they argue that the relevant phenomenology of the double slit experiment can be adequately represented by two different types of measurements with a Mach-Zehnder interferometer instead. Then they argue that the Mach-Zehnder phenomenology can be explained by a variation of Spekkens' toy model (previously laid out in \cite{Spekkens2007Evidence}).

The toy model which they use has a basis of ontic states, but a restriction on the knowledge which an observer can have of these ontic states. They called this the ``epistemic restriction'' (or "knowledge balance principle"). It basically enforces the uncertainty principle: that is, that two conjugate variables cannot be known with certainty at the same time. 

An important consequence of this epistemic restriction is that when a measurement is conducted, then the knowledge that observers have about the ontic state of the system has to be updated. This update is generically non-local -- it is Einstein's ``spooky action at a distance'' \cite{born1969albert}. This leads to the claim that Spekkens' original toy model \cite{Spekkens2007Evidence} is local and ontic, because the epistemic restriction merely affects the knowledge, not the ``real'' ontic state that the system is in.

Catani \textit{et al}'s model uses a state which has two different variables. These variables are referred to as the occupation number, and the phase. Each of these variables can take on only one of two values. The occupation number is either $0$ or $1$ (occupied or not) and the phase is either $0$ or $\pi$ (i.e. a sign of $+$ or $-$). The authors assign a phase to a path regardless of whether the occupation number is one or zero. This gives 16 possible combinations for the ontic state. (See e.g. diagram A2 in the Appendix of Catani \textit{et al}'s paper \cite{Catani2021Interference}).

The authors then introduce certain actions on this state-space that correspond to the action of the beam splitter, phase-shifter, and the which-path measurement. The which-path measurement is assumed to be non-destructive. A summary of the different experiments can be found in Fig.~4 in (the Appendix of) Catani \textit{et al}'s paper \cite{Catani2021Interference}.

The above description is all one needs to know to understand what follows. However, before we can analyse Catani \textit{et al}'s argument, we have to clarify the terminology.

\section{Ontic and Epistemic}

In Spekkens' original toy model paper \cite{Spekkens2007Evidence}, an ``ontic state'' is defined as ``a state of reality'' whereas an ``epistemic state'' is a ``state of knowledge''. 
Neither of these definitions are useful, the first because one does not know what ``reality'' means in quantum mechanics, the second because one does not know what ``knowledge'' means, whose knowledge that might be, or why we would care about it. After all, we use quantum mechanics to predict frequencies of occurrence, and not some arbitrary person's knowledge about these frequencies (note that this is not us saying anything about whether we prefer frequentist or Bayesian interpretations of classical probability, nor is it us making any claim about what quantum probabilities mean or how they are obtained---it is us taking an instrumentalist stance on what probabilities in quantum mechanics are used for). Further, these terms are defined as dichotomous, while there is nothing about the intuitive ideas of these that requires this \cite{Hance2021Wavefunctions}.

Catani \textit{et al} motivate the definitions they use for ``ontic'' and ``epistemic'' by appeal to classical physics---and specifically, to the idea of points in state space (analogous to their supposedly-ontic underlying state), and probability distributions over these (analogous to their supposedly-epistemic probability distributions, obtained from the wavefunction). They appeal to these uncontroversial classical concepts to motivate their claim that they can be applied uncontroversially in a quantum context (even in \cite{catani2022reply} straw-manning our claims that we don't know what reality or knowledge correspond to in quantum mechanics as equivalent to us claiming we cannot understand the difference between the statements ``the back door is locked'' and ``I know that the back door is locked'')---however, just because two concepts can be easily differentiated in one context, it does not mean that can in all). The debate about what counts as reality and what counts as knowledge in quantum mechanics has gone on as long as the field has existed, and claiming to be able to solve the debate by simple appeal to the classical split between these concepts when dealing with classical phase-spaces is not insightful.

To further muddy the waters, by virtue of the ``epistemic restriction'' of Spekkens' original toy model, no observer can ``know'' what the ``ontic'' state is, which makes it rather unclear what it might mean for it to be ``real'' in the first place (unlike the phase-space point in classical physics, which could in principle be measured, even if statistical mechanics deals with epistemic probability distributions over these). This inability to access underlying ``ontic'' states which the probabilistic wavefunction could be an epistemic distribution over, of course, is the discussion that Einstein and Bohr had a century ago, and the reason why the status of the wave-function is still controversial. However, the authors still write in \cite{catani2022reply} as if it is trivial to export these concepts from statistical mechanics to quantum physics.

The attempt to analogise with statistical mechanics is also problematic is, in their toy model theory, ``the incomplete knowledge is due to a fundamental feature of the physical theory... rather than a technological limitation''. While Catani \textit{et al} in \cite{catani2022reply} mock our question of whose knowledge is being given by the toy model, claiming ``statistical mechanics textbooks are right not to waste space answering the question ``whose knowledge?'' '', the question of defining the epistemic distributions given by statistical mechanics as being a representation only for those without the technology to directly manipulate the ontic state is often discussed, as ignoring it when considering the applicability of statistical mechanics leads to seemingly paradoxical, Maxwell's demon-like behaviour. Whose knowledge (or, what restrictions on our ability to access and affect the underlying ontic state are necessary for statistical mechanics to be a valid representation of the physics of a situation), is a very necessary question to ask, and a form of this question is necessary whenever a model represents some limitation on access and control over a situation as an epistemic restriction. Catani \textit{et al}'s epistemic restriction being fundamental, and so there not being an answer for this question for their model, therefore shows just how different their model is from any other case where a theory gives epistemic distributions over underlying ontic states, and so reaffirms that they need for more valid reasoning than they give for being able to export concepts from these other cases to quantum mechanics. 

The reader may find this nitpicking on words, but the words matter for the following reason. In Spekkens' original toy model paper \cite{Spekkens2007Evidence}, it is argued that the model is ``local'' because the update of the ``epistemic state'' during measurement is not an action on the (``real'', underlying) ontic state. But the authors are arguing that the model correctly reproduces the phenomenology locally. So, one needs to know what happens with the ontic state because that's what we eventually measure, and what we would expect needs to be proved to evolve and interact locally in order to say the model correctly reproduces the phenomenology locally. 

Indeed, one may wonder, why talk about knowledge at all? What we need to predict measurement outcomes is a prescription for the distribution of an ensemble of ontic states \cite{Hance2021Ensembles}. The claim of Catani \textit{et al}, that they can correctly reproduce observations, only make sense if the ``epistemic restriction'' is a change to the underlying distribution of ontic states. 

This, however, brings up the problem that, if the update due to the epistemic restriction has to affect the ensemble of ontic states, then it should only act locally, in contrast to what was previously stated in Spekkens' original toy model paper \cite{Spekkens2007Evidence}. This is why Catani \textit{et al} have to make their model more complicated.

Even worse, Catani \textit{et al}'s lengthy elaborations about epistemic states are ultimately superfluous, because in the end we still need a prescription to calculate frequencies of measurement outcomes, and that prescription is one which is equivalent to injecting a random disturbance to the ensemble of ontic states in certain cases. Given we would expect the effect of changes to our knowledge not to have a physical effect on the ontic states, and so not to be demonstrably equivalent to changing the underlying states, this provides yet another argument against considering their epistemic distribution as solely relating to knowledge. We discuss Catani \textit{et al}'s prescription in Section~\ref{sect:MI}. (For more issues with Spekkens' ontological models framework, and attempt to map these statistical mechanics concepts to quantum mechanics, see \cite{Oldofredi2020Classification,Hance2021Wavefunctions,Hance2021Ensembles}.

\section{The Model violates Measurement Independence}\label{sect:MI}

As mentioned, Catani \textit{et al} show in the Appendix that their ``knowledge balance principle'' is equivalent to an update of the ensemble of underlying states. Specifically, in Appendix C2 and C3, they provide an explanation for how the ensemble of states is supposed to be updated on beamsplitters and measurement devices

In this Appendix, the authors explain that a which-path measurement induces a random disturbance on the phase of the mode, even if the occupation number of that mode is 0. This is the case both for an interaction-free and a destructive measurement. One issue is that a measurement which changes the mode can't be interaction-free, but we will leave this aside because this isn't all that relevant and could probably be avoided.

The main problem with their model is that the authors inject some ``randomness'' into the evolution of the modes, depending on what measurement is being made (while at the same time stipulating that the model is deterministic, we may note). That is, their model is a hidden variables model, where the hidden variable is the random input. And this input depends on the measurement that is being made, because no random input comes into play for phase-measurements.

This means the model violates measurement independence. Such models are often referred to as superdeterministic \cite{Hossenfelder2020Rethinking} and they are the only known locally causal way to complete quantum mechanics. 

It is easy to see that by injecting the right amount of randomness in suitable locations into the time-evolution of a system (or its modes or whatever you want to call them) one can reproduce the outcomes of the double-slit experiment or indeed any other experiment, locally. This shouldn't be surprising because it is well known that by violating measurement independence the statistical predictions of quantum mechanics can be reproduced while maintaining local causality. 

The authors even mention in \cite{catani2022reply} that ``for the case of a measurement on one mode, the ensemble of possibilities for the physical state of the other mode may be updated \textit{because of a pre-existing correlation between the physical states of the two modes}'', and use this to argue that therefore, such updating does not involve any nonlocal influence. This is correct, but in this case this pre-existing correlation must be dependent on what choice of measurement is made---and so, violate measurement independence!

Possibly the reason the authors didn't notice they violate measurement independence is that it is often implicitly assumed to require a distribution of hidden variables at the time of preparation that must depend on the measurement settings. However, as we pointed out in \cite{Hance2021Ensembles}, this is not correct. It is instead sufficient if the {\emph{evolution}} of the ontic states depends on what is being measured. One way to do this is by injecting suitable amounts of randomness on paths with beamsplitters or measurement devices that made no detections. 
Catani \textit{et al} have therefore put forward a further example of what we called a $\psi$-ensemble model.

Reproducing quantum mechanics by injecting randomness locally is particularly simple in the case that Catani \textit{et al} consider because in the Mach-Zehnder interferometer the paths recombine before the measurement. This is not the case for the double-slit experiment, so their construction cannot be readily used for the case that Feynman was actually concerned with. However, if one allows a measurement-dependent injection of randomness, one can also reproduce the double-slit, though the model would have to be more difficult.

Since their model is arguably not fine-tuned, it also provides another example that superdeterministic hidden variables theories don't have to be fine-tuned, contrary to the claim that was made in \cite{Daley2022Adjudicating} (and countered in \cite{Hance2024DaleyComment}). Another counterexample of the claim provided in \cite{Donadi2022SuperdetToy}.

While using measurement dependence to reproduce this effect is local, it is far more debatable that it is classical, given measurement dependence allows us to reproduce all quantum mechanical effects. Therefore, basing their argument that interference phenomena can be reproduced classically on a measurement-dependent model does not support their claim that such phenomena are not linked to the mysteries of quantum mechanics.

\section{Conclusion}

We have argued here that Catani \textit{et al} fall short of their goal in showing that interference phenomena do not capture the essence of quantum theory. They only succeed in reproducing the phenomenology of the Elitzur-Vaidman bomb tester by providing a toy model for which measurement independence is violated. Given measurement independence violation can be used to reproduce all quantum-mechanical phenomena, using such a model to reproduce interference phenomena if anything reinforces the link between interference phenomena and quantum phenomena.

We appreciate the development of new toy models as a way to link certain assumptions with certain conclusions. However, we do not think it is surprising that it is possible to locally reproduce quantum phenomena, if one's model is measurement-dependent.

\textit{Acknowledgements:}
We thank Lorenzo Catani, Matt Leifer, David Schmid, and Rob Spekkens for their patience.
SH acknowledges support by the Deutsche Forschungsgemeinschaft (DFG, German Research Foundation) under grant number HO 2601/8-1. JRH is supported by Hiroshima University's Phoenix Postdoctoral Fellowship for Research,  the University of York's EPSRC DTP grant EP/R513386/1, and the EPSRC Quantum Communications Hub (funded by the EPSRC grant EP/M013472/1).

\bibliographystyle{unsrtnat}
\bibliography{ref.bib}

\appendix

\section{Classicality}

Catani \textit{et al} state in their paper that ``Whatever principle of classicality one adopts, it is clear that the toy field theory should come out as a classical theory by its lights'' \cite[p21 (V. Discussion: A. Take Away: 3. Beyond the TRAP phenomenology: what is really nonclassical?)]{Catani2021Interference}. While their definition of classicality should therefore not affect their argument (and so our rebuttal above), they clarify that in their own view classicality amounts to: 

\begin{enumerate}
    \item Physical states are represented by a set while dynamics are represented by functions over this set;
    \item\label{classp2} Inferences are done using Bayesian probability theory and Boolean propositional logic; and,
    \item That which is empirically indiscernible should be ontologically identical.
\end{enumerate}

First, it seems rather odd to us to require the use of Bayesian probability theory which for the most part isn't used much in (what is commonly called) classical statistical mechanics. It seems that according to Catani \textit{et al}'s terminology, this would not be classical. This however seems to be as, by point \ref{classp2}, rather than inferences requiring/being done using Bayesian probability theory and Boolean propositional logic, they actually mean that inferences \textit{do not contradict} what the inferences would be were they done using Bayesian probability theory and Boolean propositional logic. That they mistook `require' with `do not contradict' seems implicit in their response in \cite{catani2022reply} to us criticising their point \ref{classp2}, and makes far more sense.

Second, what is empirically `indiscernible' depends on what measurements one one has made or can make. Distances below, say, a thousandth of a femtometre aren't currently `empirically discernible'. We treat them as ontologically different in General Relativity, hence, it seems that according to the authors' position, General Relativity is not a classical theory. To circumvent this point, Catani \textit{et al} in \cite{catani2022reply} claim that the form of empirical indiscernibility they consider is one ``in principle'' rather than one ``in practise'' (i.e. one limited by technology). However, what is empirically discernible in principle, beyond what has already been shown to be empirically discernible in practise by experiment, is inherently theory-dependent, which makes this a bad measure to use when trying to determine classicality.

Third, in absence of a clear definition for what `ontological' even means, it seems that according to this definition any theory can be turned into a quantum theory just by postulating the existence of a single ontological state that can't be empirically discerned.

Now, a definition is just a definition, but a definition that is so in conflict with how a technical term is currently used is not one that we think should widely be adopted. 

\end{document}